\documentclass[useAMS,usenatbib]{mn2e}
\usepackage{amsmath,amssymb,graphicx,color,longtable}
\title {The effect of macromodel uncertainties on microlensing modelling of lensed quasars}
\author[G. Vernardos, C.J. Fluke]
  {G. Vernardos$^1$\thanks{gvernard@astro.swin.edu.au}, C.J. Fluke$^1$\\
  $^1$Centre for Astrophysics \& Supercomputing, Swinburne University of Technology, PO Box 218, Hawthorn, Victoria, 3122, Australia\\}
\pagerange{\pageref{firstpage}--\pageref{lastpage}}
\pubyear{2014}

\begin{document}
\label{firstpage}
\maketitle

\begin{abstract}
Cosmological gravitational microlensing has been proven to be a powerful tool to constrain the structure of multiply imaged quasars, especially the accretion disc and central supermassive black--hole system.
However, the derived constraints on models may be affected by large systematic errors introduced in the various stages of modelling, namely, the macromodels, the microlensing magnification maps, and the convolution with realistic disc profiles.
In particular, it has been known that different macromodels of the galaxy lens that fit the observations equally well, can lead to different values of convergence, $\kappa$, and shear, $\gamma$, required to generate magnification maps.
So far, $\sim$25 microlensed quasars have been studied using microlensing techniques, where each system has been modelled and analyzed individually, or in small samples.
This is about to change due to the upcoming synoptic all-sky surveys, which are expected to discover thousands of quasars suitable for microlensing studies.
In the present study we investigate the connection between macromodels of the galaxy lens and microlensing magnification maps throughout the parameter space in preparation for future studies of large statistical samples of systems displaying microlensing.
In particular, we use 55,900 maps produced by the GERLUMPH parameter survey (available online at {\tt http://gerlumph.swin.edu.au}) and identify regions of parameter space where macromodel uncertainties ($\Delta\kappa,\Delta\gamma$) lead to statistically different magnification maps.
Strategies for mitigating the effect of $\Delta\kappa,\Delta\gamma$ uncertainties are discussed in order to understand and control this potential source of systematic errors in accretion disc constraints derived from microlensing.
\end{abstract}


\section{Introduction}
\label{sec:intro}
Gravitational microlensing on cosmological scales is a powerful tool for studying the structure of quasars with unprecedented detail \citep[see][for a review]{Schmidt2010}.
The effect of stellar mass objects near the line of sight of multiply imaged quasars is sensitive to scales from the central accretion disc and X--ray emitting region \citep[$\sim 10^{14}$ cm, e.g.][]{Poindexter2008,Dai2010} to the broad emission--line region \citep[$\sim 10^{17}$ cm, e.g.][]{ODowd2011,Sluse2012}.
However, the derived constraints on accretion disc models, in particular with respect to the thin-disc model \citep[][]{Shakura1973}, have been found to be marginally consistent \citep[e.g.][]{Eigenbrod2008,Poindexter2008,Anguita2008,Bate2008}, or in disagreement with model predictions \citep[][]{Floyd2009}.

From the sample of $\sim$90 known multiply imaged quasars \citep{Mosquera2011b}, only $\sim$25 systems have been studied in detail using microlensing techniques \citep{Bate2012}.
This is mostly due to the difficulty of the related observations, requiring either long-term monitoring \citep[the light--curve analysis method, see][for examples and applications]{Kochanek2004,Morgan2010,Mosquera2013}, or simultaneous multi-wavelength observations \citep[the snapshot method, e.g.][]{Bate2008,Floyd2009,Blackburne2011,Jimenez2014}.
Nevertheless, the upcoming synoptic all--sky survey facilities, such as the Large Synoptic Survey Telescope \citep[LSST;][]{LSST2009}, PanSTARRS \cite[][]{Kaiser2002}, and SkyMapper \citep[][]{Keller2007}, are estimated to discover thousands of multiply imaged quasars \citep[][]{Oguri2010} and provide observations suitable for microlensing studies.

It is timely and crucial to understand the origin of systematic uncertainties in the microlensing-derived accretion disc model constraints that are introduced in the various stages of the modelling process.
Quasar microlensing models can be broken down to three distinct components:

\begin{enumerate}
\item The mass distribution of the galaxy lens viz. the macromodel.
The gravitational lens equation is solved for different mass distributions, either analytically \citep[e.g. see][]{Witt1995}, or more often numerically \citep[e.g. the {\tt GRAVLENS} software\footnote{http://redfive.rutgers.edu/$\sim$keeton/gravlens/};][]{Keeton2001}.
The best-fitting model is selected, which reproduces a number of observables such as the positions of the multiple images, their relative fluxes, the time delays (if available), etc.

\item The microlensing magnification map.
This map is a pixellated version of the source plane, approximating the magnification patterns, or caustics, produced by the compact microlenses \citep[e.g.][]{Kayser1986}.
Generating a map requires parameters derived from the preceding macromodelling stage, namely, the convergence, $\kappa$, describing the surface mass density in the lens plane, and the shear, $\gamma$, describing the tidal shear field.
An additional parameter is the smooth matter fraction, $s=\kappa_{\rm s}/\kappa$, which measures the contribution from smoothly distributed matter ($\kappa_{\rm s}$) and compact microlenses ($\kappa_{*}$)to convergence, i.e. $\kappa = \kappa_{\rm s} + \kappa_{*}$.

\item A model profile for the source e.g. an accretion disc profile.
The magnification map is assumed to represent the microlensing effect on a point source.
To model the effects on extended sources, the map must be convolved with a source profile \citep[see][for different examples of this process]{Kayser1986,Kochanek2004,Bate2008}.
\end{enumerate}

\cite{Vernardos2013} investigated a possible source of systematic errors in magnification maps, originating from different placement of the microlenses and leading to caustic configurations with different statistical properties.
They found that in specific regions of parameter spaces there is a $\sim 7$ per cent probability of producing a map with a different magnification probability distribution (MPD).
\cite{Mortonson2005} examined a list of simplified shapes of the accretion disc profile, concluding that the MPDs of convolved maps are relatively insensitive to all properties of the models except the half-light radius of the disc.

For the majority of published macromodels, the best-fitting $\kappa,\gamma$ values are quoted without uncertainties \citep[a rare counter--example is][]{Schmidt1998}.
However, it is recognised that there could be alternative macromodels which could fit the observational data equally well \citep[e.g.][]{Mediavilla2009}.
Uncertainties in the derived $\kappa,\gamma$ values can be taken into account indirectly by allowing an uncertainty in the resulting macro--image magnifications:
\begin{equation}
\label{eq:mu_th}
\mu_{\rm th} = \frac{1}{(1-\kappa)^2-\gamma^2} \, .
\end{equation}
For example, a 10 per cent variation in the macro--magnification can lead to uncertainties $0.001 < \Delta\kappa,\Delta\gamma < 0.03$, depending on the actual location in the $\kappa,\gamma$ parameter space.
This approach has been followed in \cite{Morgan2006} and \cite{Blackburne2011}, where systematic uncertainties of $\le$0.05 mag have been added to the observed flux ratios between multiple quasar images.

It is known that the density distribution in the central regions of galaxies should be cuspy \citep[e.g.][]{Faber1997}, with the isothermal spherical mass distribution being a realistic representation \citep[e.g.][]{Fabbiano1989,Kochanek1995,Rix1997}.
\cite{Mediavilla2009} modelled the lensing galaxies in 20 systems as singular isothermal spheres (SIS) with external shear \citep[e.g. see][]{Witt1995}, providing the $\kappa,\gamma$ extracted from their models as compatible values with high uncertainties.
\cite{Blackburne2011} used the same modelling approach for 12 systems, choosing to increase the complexity of the SIS model.
Where a poor $\chi^2$ goodness of fit to the data was present, they used an isothermal ellipsoid (for SDSS J$1330+1810$), added a second isothermal sphere to account for satellite galaxies of the galaxy lens (for HE $0230-2130$, MG $0414+0534$, RX J$0911+0551$, and WFI J$2033-4723$), or, in case the galaxy lens was in fact part of a group of galaxies, added a second isothermal sphere to account for the rest of the group members (PG $1115+080$).

Another macromodel approach is to use a de Vaucouleurs profile tracing the light distribution of the galaxy lens (stars; compact matter component), which is embedded in a dark matter halo (smooth matter component), with external shear \citep[][]{Lehar2000,Morgan2006,Morgan2008,Dai2010}.
The advantage of this technique is that it provides a constraint on the fraction of compact to smooth matter, which cannot be constrained by the SIS models.
Nevertheless, the uncertainty in $\kappa,\gamma$ persists, and a series of models fit equally well the observed properties \citep[e.g.][]{Morgan2006}.

Finally, macromodels for any specific system can be effected by the mass--sheet degeneracy \citep[][]{Falco1985,Gorenstein1988}.
Transfoming the macromodel of the galaxy lens by scaling the mass distribution and adding a constant surface mass density (mass--sheet) leaves observables such as the image positions, shapes, fluxes, etc, unchanged.
Therefore, more information about the source (e.g. absolute luminosity or size) or the lens (e.g. mass derived from observations of stellar dynamics) is required to uniquely constrain the mass distribution of the lens, and consequently the $\kappa,\gamma$ values.

Taking for granted that there will be uncertainties in the $\kappa,\gamma$ parameters, we investigate in this study the connection between such macromodel uncertainties and the resulting microlensing magnification maps, throughout the $\kappa,\gamma,s$ parameter space.
In Section 2 we describe the use of 55,900 microlensing magnification maps from the Graphics Processing Unit--Enabled High Resolution MicroLensing parameter survey \citep[GERLUMPH;][]{Vernardos2014a,Vernardos2014b} to achieve our goal.
Our results comparing the magnification maps in terms of their magnification probability distribution throughout the parameter space are presented in Section 3.
Discussion on the microlensing, macromodelling, and accretion disc implications of our results follows in Section 4.
We present our conclusions in Section 5.

\begin{figure*}
\includegraphics[width=\textwidth]{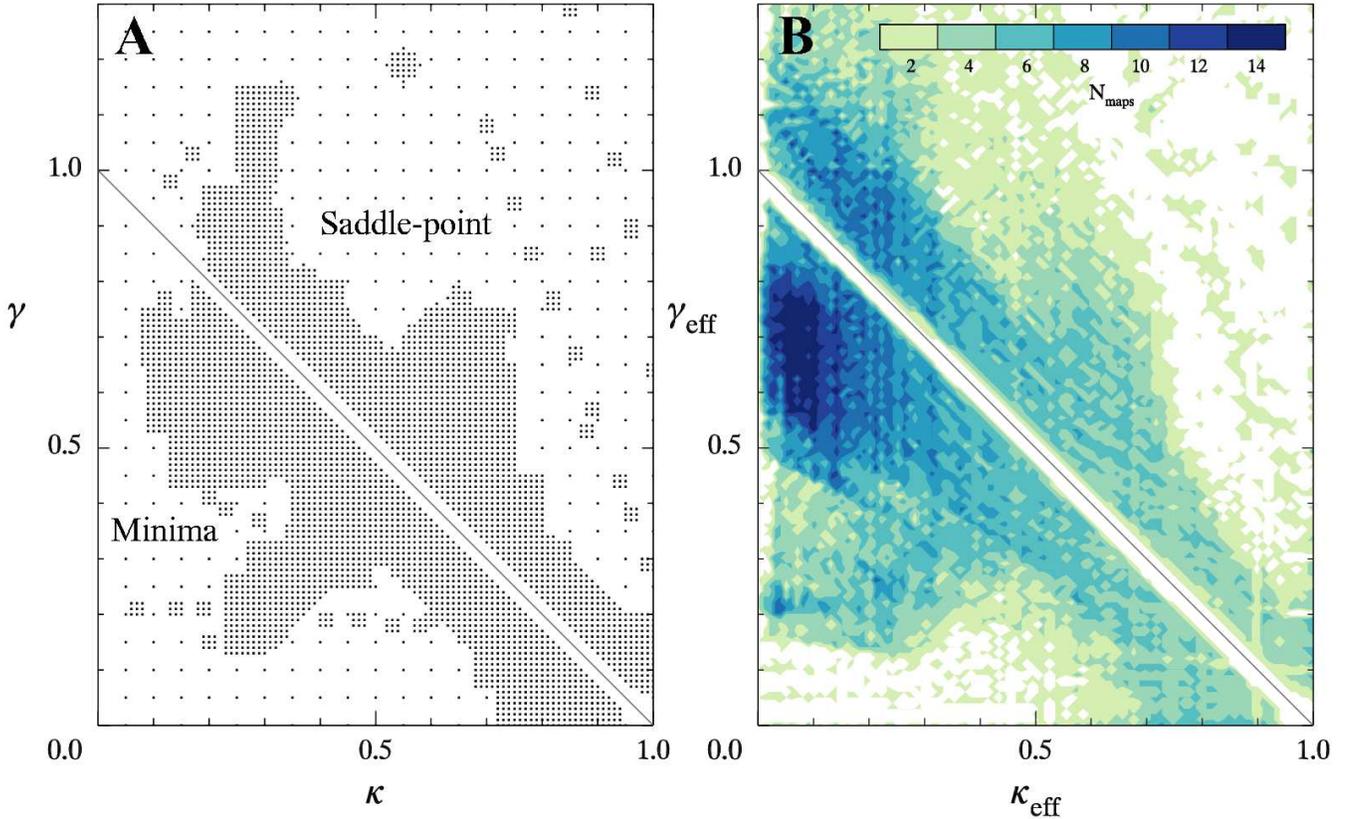}
\caption{Panel A: GERLUMPH coverage of the $\kappa,\gamma$ parameter space. Maps have been generated on an extended grid of 0.05 separation, $\pm 0.01$ from macromodel values found in the literature, and on an extended area uniformly covered by a grid of 0.01 separation. At each location we use 10 maps with $0.0 \leq s \leq 0.9$ in steps of 0.1. Panel B: GERLUMPH coverage of the $\kappa_{\rm eff},\gamma_{\rm eff}$ parameter space, computed from the $\kappa,\gamma$ shown in panel A and all values of $s$ using equation (\ref{eq:eff}). The $\kappa_{\rm eff},\gamma_{\rm eff}$ space is not uniformly covered, with the densest regions appearing as we move away on straight lines radiating from (1,0).\label{fig:pspaces}}
\end{figure*}

\begin{figure}
\begin{center}
\includegraphics[width=0.3\textwidth]{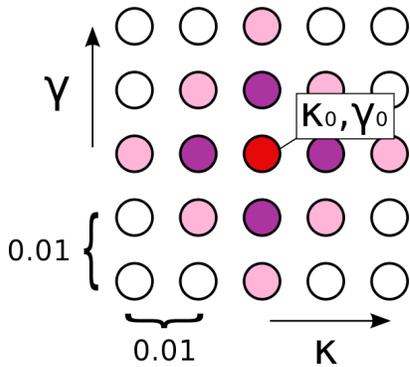}
\caption{Schematic representation of a $\kappa,\gamma$ area in parameter space which is uniformly covered by maps. As the radius $r$ of a circular area centered at $\kappa_0,\gamma_0$ (red circle) increases, the MPDs of more neighbouring maps are compared with each other. For $r=0.01$ there are 4 neighbouring maps included (purple circles), for $r=0.02$ there are 8 (pink circles), and similarly for higher values of $r$ not shown here. This process is repeated for all values of the parameter $s$.\label{fig:schematic}}
\end{center}
\end{figure}

\begin{figure*}
\includegraphics[width=\textwidth]{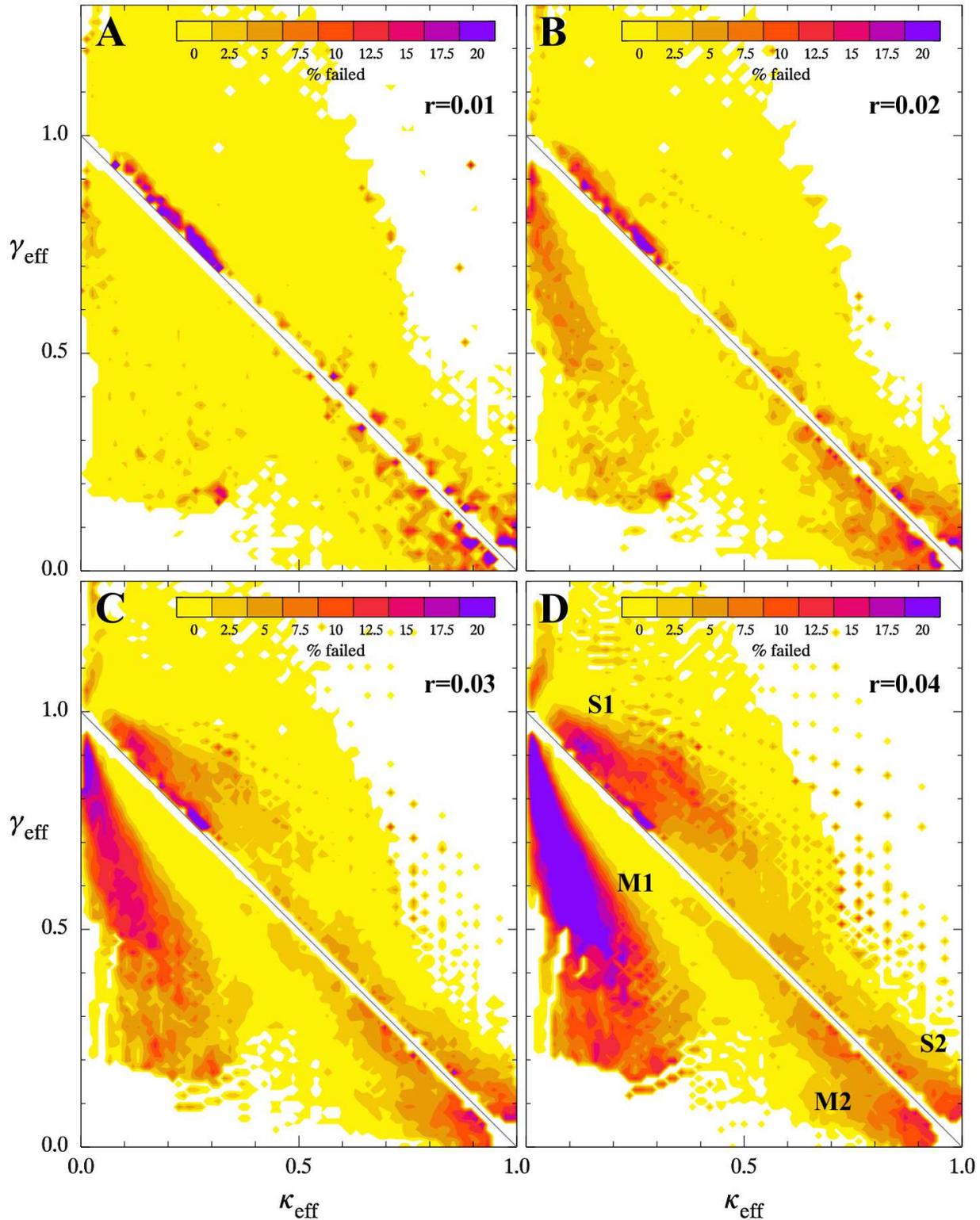}
\caption{Percentage of failed Kolmogorov--Smirnov tests between all possible pairs of magnification probability distributions from maps within a circular area of $r=$ 0.01, 0.02, 0.03, and 0.04, in the $\kappa_{\rm eff},\gamma_{\rm eff}$ parameter space. Areas with increased numbers of failed tests appear for $r \ge 0.02$, labelled as M1, M2 in the minima, and S1, S2 in the saddle--point regions in panel D.\label{fig:eff_pairs}}
\end{figure*}

\section{Method}
GERLUMPH\footnote{\tt http://gerlumph.swin.edu.au} is an open resource of simulated microlensing data, currently consisting of $>$70,000 magnification maps, complemented by online analysis tools \citep[][]{Vernardos2014b}.
The magnification maps were produced using the {\tt GPU-D} direct inverse ray--shooting technique \citep[][]{Thompson2010,Thompson2014,Vernardos2014b}, on the GPU--Supercomputer for Theoretical Astrophysics Research (gSTAR).

We use the GERLUMPH maps that are located in the range $0.0 < \kappa \le 1.0$ and $0.0 \le \gamma \le 1.3$, covering large areas uniformly with $\Delta\kappa,\Delta\gamma = 0.01$, with 10 values of $s$ between 0.0 and 0.9 in steps of 0.1 for each $\kappa,\gamma$ combination.
This set consists of 5590 $\kappa,\gamma$ locations (multiplied by 10 for the different values of $s$) shown in panel A of Fig. \ref{fig:pspaces}; this is where most of the macromodels for existing systems are located \citep[e.g.][]{Bate2012}.
The critical line, i.e. where $\mu_{\rm th} \rightarrow \infty$ from equation (\ref{eq:mu_th}), divides the parameter space into the minima ($1-\kappa-\gamma > 0$, below the grey line on Fig. \ref{fig:eff_pairs}) and saddle--point ($1-\kappa-\gamma < 0$, above the grey line on Fig. \ref{fig:eff_pairs}) regions.
These regions of parameter space correspond to the extrema of the light--travel surface, where the macro--images of the background quasar form \citep[e.g. see][]{Blandford1986}.
For the GERLUMPH maps, the map width is set to 25 $R_{\rm Ein}$, where
\begin{equation}
\label{eq:Rein}
R_{\rm Ein} = \sqrt{ \frac{D_{\rm os}D_{\rm ls}}{D_{\rm ol}} \frac{4G\langle M \rangle}{c^2} } \, .
\end{equation}
is the Einstein radius of the gravitational lens system, with $D_{\rm ol}$, $D_{\rm os}$, and $D_{\rm ls}$ being the angular diameter distances from observer to lens, observer to source, and lens to source, and $\langle M \rangle$ is the mean mass of the microlenses.
The map resolution is 10000 pixels per dimension, the microlenses are distributed randomly over the lens plane and the microlens masses are all 1 M$_{\odot}$.

The parameters of each map are stored and managed by a relational database, facilitating access to the map data.
The map MPD, i.e. the probability distribution of the magnification values in the map pixels, has been precomputed and stored alongside the actual map data.
We access the MPDs through the GERLUMPH database and use them to produce our results in Section \ref{sec:results}.

\subsection{Parametrization}
As described in Section \ref{sec:intro}, certain macromodel approaches can constrain all three of the $\kappa,\gamma,s$ parameters, while others constrain only the $\kappa,\gamma$ and leave $s$ to be treated as a free parameter for generating magnification maps.

As \cite{Paczynski1986} has shown, the three-dimensional $\kappa,\gamma,s$ parameter space is equivalent to the two-dimensional effective parameter space, $\kappa_{\rm eff},\gamma_{\rm eff}$:
\begin{equation}
\label{eq:eff}
\kappa_{\rm eff} = \frac{(1-s)\kappa}{1-s\kappa} \, , \, \gamma_{\rm eff} = \frac{\gamma}{1-s\kappa}
\end{equation}
where $\kappa_{\rm eff}$ is entirely due to compact matter.

The advantage of using the transformation of equation (\ref{eq:eff}) is that it allows for properties calculated in three dimensions ($\kappa,\gamma,s$) to be displayed in just two ($\kappa_{\rm eff},\gamma_{\rm eff}$, see Fig. \ref{fig:pspaces}).
On the other hand, uncertainties in the $\kappa_{\rm eff},\gamma_{\rm eff}$ imply an additional $\Delta s$, along with the $\Delta\kappa,\Delta\gamma$.
With respect to the GERLUMPH maps, the $\kappa,\gamma,s$ parameter space is covered uniformly (panel A of Fig. \ref{fig:pspaces}), while the $\kappa_{\rm eff},\gamma_{\rm eff}$ space is covered more densely as we move outwards along locii radiating from (1,0) (panel B of Fig. \ref{fig:pspaces}).

In the following, we have used the $\kappa,\gamma,s$ parametrization to produce our results and both parametrizations to display them.
We add comments where necessary to clarify the interpretation of our results.

\subsection{Hypothesis testing -- the KS test}
One way to compare magnification maps is through their MPDs.
Basic statistical properties like the mean, median, mode, skewness, etc, can be extracted, or more advanced statistical comparisons, such as $\chi^2$ tests \citep[e.g.][]{Mortonson2005} or Kolmogorov--Smirnov (KS) tests \citep[e.g.][]{Vernardos2013,Vernardos2014a} can be used.

In this work, we use the KS test to compare MPDs of neighbouring maps throughout the parameter space.
This test calculates the maximum absolute difference, $D$, between the cumulative probabilities of the two distributions being tested (two--sided two--sample KS test).
Based on this measurement and the sample size, a p--value can be returned from the function:
\begin{equation}
\label{eq:Qks}
Q( s )= 2 \sum_{j=1}^{\infty} (-1)^{j-1} e^{-2 j^2 s^2} \, ,
\end{equation}
which is a monotonic function with the limiting values:
\begin{equation}
\label{eq:limits}
\lim_{s \to 0} Q( s )= 1 \quad {\rm and} \quad \lim_{s \to \infty} Q( s )= 0 \, .
\end{equation}
For the variable $s$ we use the approximation of \cite{Stephens1970}:
\begin{equation}
\label{eq:approx}
s = \left ( \sqrt{N} + 0.12 + \frac{0.11}{\sqrt{N}} \right ) D \, ,
\end{equation}
where $N=N_{\rm A} N_{\rm B} / ( N_{\rm A} + N_{\rm B})$ and $N_{\rm A}$,$N_{\rm B}$ are the sizes of the two sampled distributions.
This approximation becomes asymptotically accurate for large values of $N$.
More details on the KS test can be found in \cite{Hollander1973} and \cite{Durbin1973}.

We investigate the null hypothesis that the MPDs are statistically equivalent, which is rejected whenever a p--value of $< 5$ per cent is found (the test fails).
However, we are interested in observing trends of the test results throughout the parameter space, not in the actual results themselves.
This means that we can allow for reduced precision for our calculated p--values.
The MPDs of the GERLUMPH maps consist typically of $\mathcal{O}(10^4)$ magnification values \citep[see table B.5 of][]{Vernardos2014b}.
Although the KS test should be used with uncensored and ungrouped data, we choose to group our MPDs to 100 bins \cite[e.g.][]{Vernardos2013} within the magnification range of interest.
In this way, $N$ is dramatically decreased while $D$ stays roughly the same, and the p--values returned are artificially inflated (equation \ref{eq:limits}).
Therefore, binning the distributions is a somewhat conservative approach regarding the actual KS test results, but it is sufficient to examine trends in the parameter space.

\begin{figure}
\includegraphics[width=0.47\textwidth]{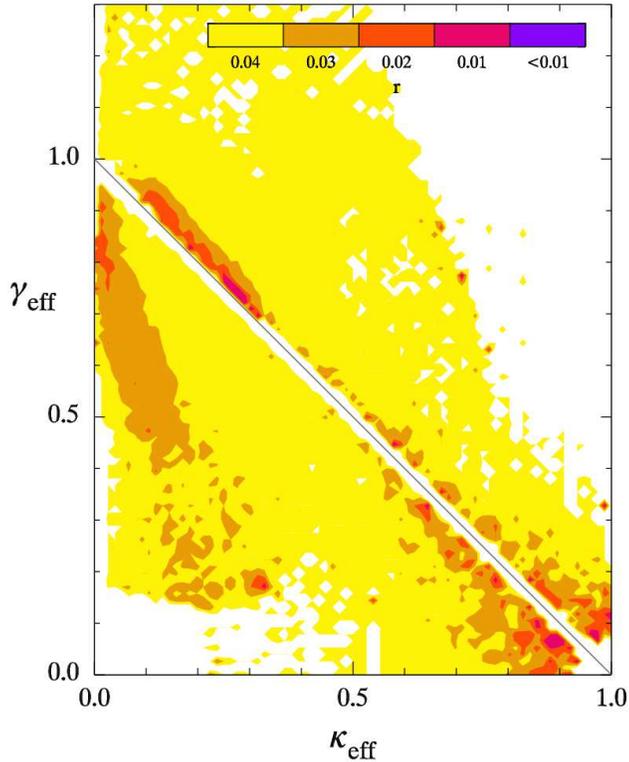}
\caption{Value of $r$ within which $<$7 per cent of neighbouring map MPDs failed the KS test with the central MPD, in the $\kappa_{\rm eff},\gamma_{\rm eff}$ parameter space. The M1, M2, S1, S2, areas appearing in panels B, C, and D of Fig. \ref{fig:eff_pairs} can be also seen here.\label{fig:eff_k0g0}}
\end{figure}

\section{Results}
\label{sec:results}
We use two different methods to perform the KS test: among all the possible pairs of MPDs located within a given macromodel uncertainty radius and between the MPDs of a fiducial macromodel and all available neighbouring maps (see Fig. \ref{fig:schematic}).
In the first case, all maps within a given area of uncertainty in the $\kappa,\gamma$ parameter space are treated equally and compared to each other, while in the second case a fiducial $\kappa_0,\gamma_0$ (e.g. an existing macromodel value without uncertainty) is compared to all its neighbours (but the neighbours are not compared amongst themselves).

We compare the MPDs of all the maps located within a radius $r =$ 0.01, 0.02, 0.03, and 0.04 from every $\kappa_0,\gamma_0$ pair located throughout the parameter space of interest.
The radius $r$ can be thought of as a measure of the $\Delta\kappa,\Delta,\gamma$ uncertainty of macromodel-derived values.
By selecting the $\kappa_0,\gamma_0$ values to match existing GERLUMPH maps from the continuous $\Delta\kappa,\Delta\gamma = 0.01$ regions in parameter space (see first panel of Fig. \ref{fig:pspaces}) we end up with 5, 13, 29, and 49 maps for each value of $r$.
This is shown schematically in Fig. \ref{fig:schematic}.
We compare the corresponding MPDs by performing KS tests between 10, 78, 406, and 1176 pairs of MPDs respectively, counting the percentage of pairs that failed the test.
This process is repeated for $0.0 \leq s \leq 0.9$, with $\Delta s = 0.1$.

In Fig. \ref{fig:eff_pairs}, we show the percentage of pairs that failed the test for $r =$ 0.01, 0.02, 0.03, and 0.04 in the $\kappa_{\rm eff},\gamma_{\rm eff}$ parameter space.
We point out that although this includes the KS results for all values of $s$, it only encodes information on $\Delta\kappa,\Delta\gamma$ and no conclusions can be drawn on any effect of $\Delta s$.
Looking at the panels of Fig. \ref{fig:eff_pairs} for all values of $r$ we can see that there are large areas of parameter space where all the MPD pairs successfully pass the KS test.
For $r > 0.01$ however, areas containing increasing numbers of failing pairs appear.
Two distinct such areas appear in each of the minima and saddle--point regions; M1, M2 and S1, S2 in panel D of Fig. \ref{fig:eff_pairs}.

Next, we investigate how far we can go from a fiducial macromodel $\kappa_0,\gamma_0$ before there are significant differences between the MPDs of this macromodel and its neighbours.
The null hypothesis in this case is that the MPDs of all the maps located within a radius $r$ in the $\kappa,\gamma$ plane are equivalent to the central MPD.
As the value of $r$ increases, we compare 4, 12, 28, and 48 pairs of neighbouring MPDs with the central one (see Fig. \ref{fig:schematic}) and count the percentage of pairs that failed the test.
This process is repeated for $0.0 \leq s \leq 0.9$, with $\Delta s = 0.1$.
In Fig. \ref{fig:eff_k0g0} we show the value of $r$ at which 7 per cent of neighbouring macromodels fail the KS test, plotted in the $\kappa_{\rm eff},\gamma_{\rm eff}$ parameter space.
This value was chosen in order to prevent lens position systematics from contaminating our results (see Section \ref{sec:discussion}).

\begin{figure*}
\includegraphics[width=0.89\textwidth]{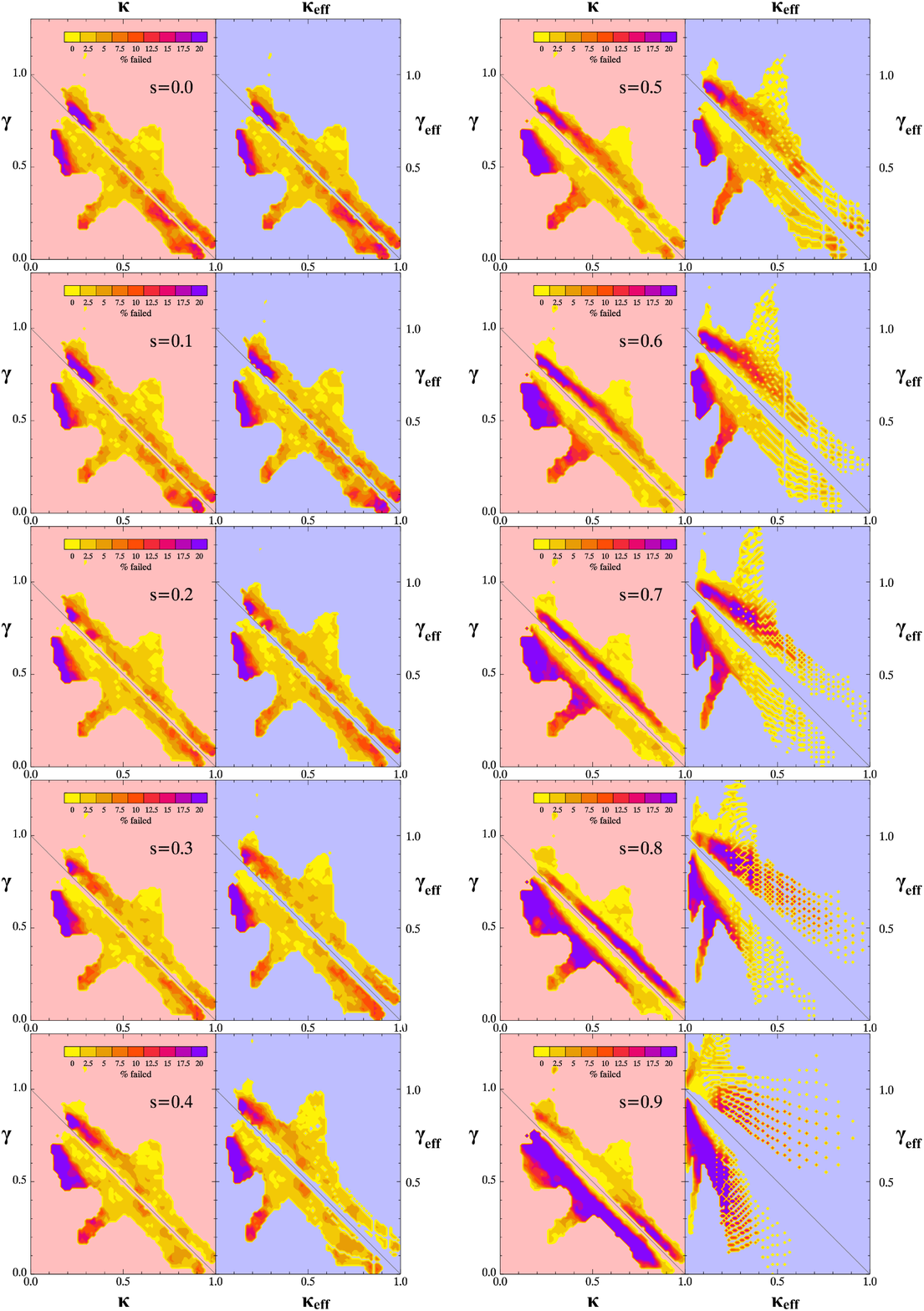}
\caption{Same information as panel D of Fig. \ref{fig:eff_pairs} plotted in the $\kappa,\gamma,s$ and $\kappa_{\rm eff},\gamma_{\rm eff}$ parameter spaces. Different values of $s$ are shown in each panel.\label{fig:kgs_pairs}}
\end{figure*}

\begin{figure*}
\includegraphics[width=0.89\textwidth]{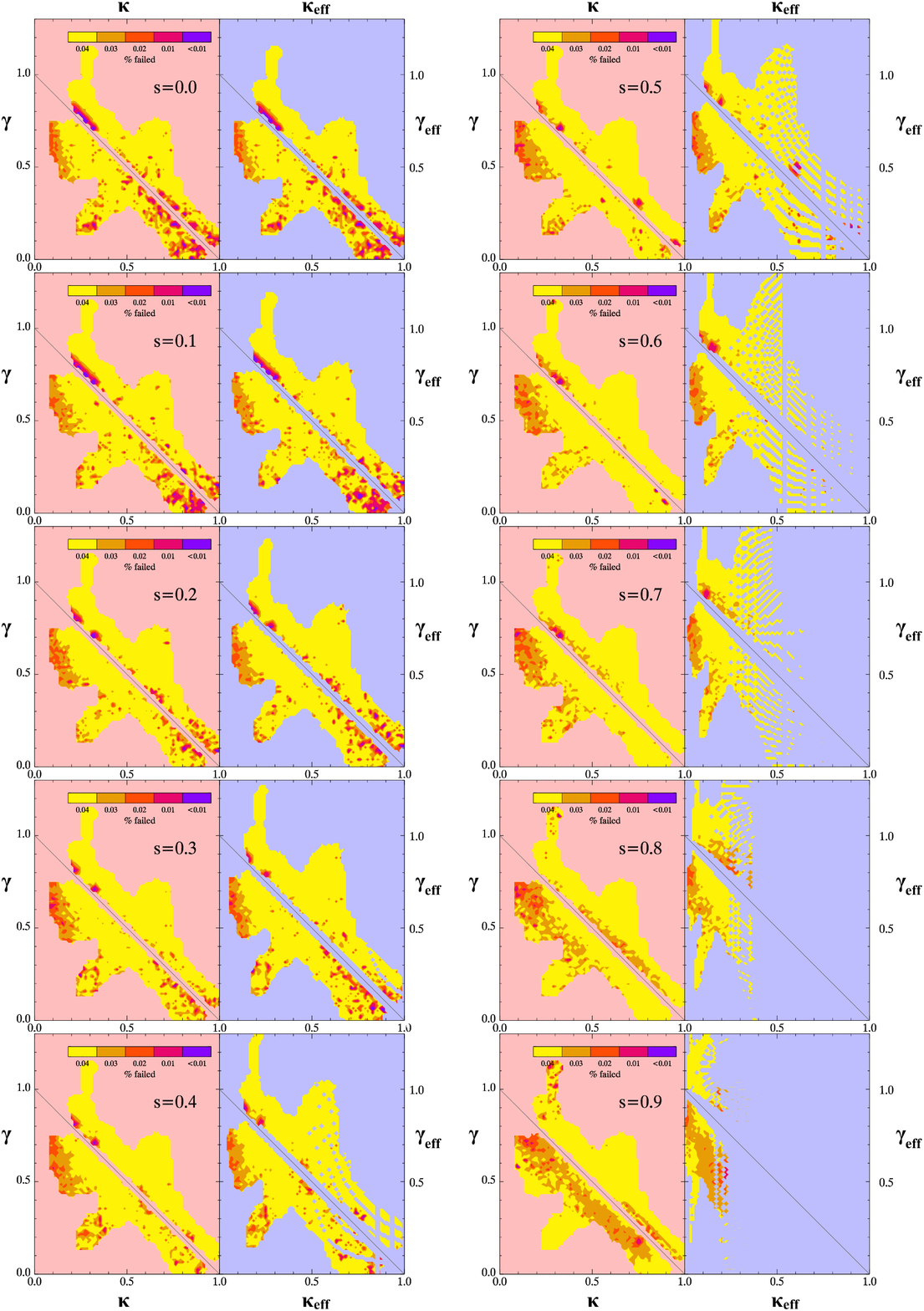}
\caption{Same information as Fig. \ref{fig:eff_k0g0} plotted in the $\kappa,\gamma$ parameter space. Different values of $s$ are shown in each panel.\label{fig:kgs_k0g0}}
\end{figure*}

Finally, in Fig. \ref{fig:kgs_pairs} and \ref{fig:kgs_k0g0} we show the same results as above, in both $\kappa,\gamma$ and $\kappa_{\rm eff},\gamma_{\rm eff}$ parameter spaces, for different values of $s$ in each panel.
Averaging between the plots in $\kappa_{\rm eff},\gamma_{\rm eff}$ for all values of $s$ produces Fig. \ref{fig:eff_k0g0} and panel D of Fig. \ref{fig:eff_pairs}.

\section{Discussion}
\label{sec:discussion}
The appearance of the parameter space in Fig. \ref{fig:eff_pairs} and \ref{fig:eff_k0g0} is almost identical, and independent of the density of maps (see panel B of Fig. \ref{fig:pspaces}), which indicates that the underlying parameter space properties remain the same regardless of which way the MPDs are compared.
Although the appearance of the $\kappa,\gamma$ parameter space in Fig. \ref{fig:kgs_pairs} and \ref{fig:kgs_k0g0} changes with respect to the value of $s$, this is not true for the $\kappa_{\rm eff},\gamma_{\rm eff}$ space shown in the adjacent panels.
It is easier to classify our results in terms of $\kappa_{\rm eff},\gamma_{\rm eff}$, although the two parametrizations are equivalent.

There are extended regions of parameter space where all the neighbouring maps for $r \leq 0.04$ have statistically equivalent MPDs (see Fig. \ref{fig:eff_pairs}).
This is in agreement with Fig. \ref{fig:eff_k0g0}, where $r$ assumes its highest values in the same regions.
Nevertheless, there are also regions with increased numbers of failed tests between neighbouring maps for $r \geq 0.02$, labelled as M1, M2, S1, and S2 in panel D of Fig. \ref{fig:eff_pairs}.
The properties and impact of these regions are examined further below.

The KS test results depend more strongly on the central MPD when we compare it with all its neighbours, but this has a small effect on our results.
For example, for $\kappa,\gamma$ found in areas of parameter space close to the critical line and for $\kappa \gtrsim 0.6$, any MPD has a $\sim 7$ per cent probability to be effected by microlens position systematics \citep[][]{Vernardos2013}.
If this happens to be the central MPD, a higher percentage of failed pairs would be observed.
In Fig. \ref{fig:kgs_k0g0} we see isolated individual points with low values of $r$ located in areas with generally high values of $r$, which are areas of parameter space prone to be effected by lens position systematics.
In this case, the central, or a number of neighbouring MPDs, may be effected by microlens position systematics causing more than 7 per cent of the compared pairs of MPDs to fail the KS test even for low values of $r$.
Increasing our tolerance of failed pairs to more than 7 per cent leads to higher corresponding values of $r$ but the appearance of the parameter space remains the same.

Our results in Fig. \ref{fig:eff_pairs} for $r=0.01$ may be effected by microlens position systematics as well.
For example, in panel A we are comparing 5 MPDs with each other (10 pairs of MPDs); any single MPD affected by microlens position systematics could lead to up to 40 per cent failed pairs.
For $r>0.01$, the number of MPD pairs is high enough for the effect of any single MPD to remain low.

\subsection{Examination in terms of microlensing}
\label{sec:micro}
\begin{figure*}
\includegraphics[width=\textwidth]{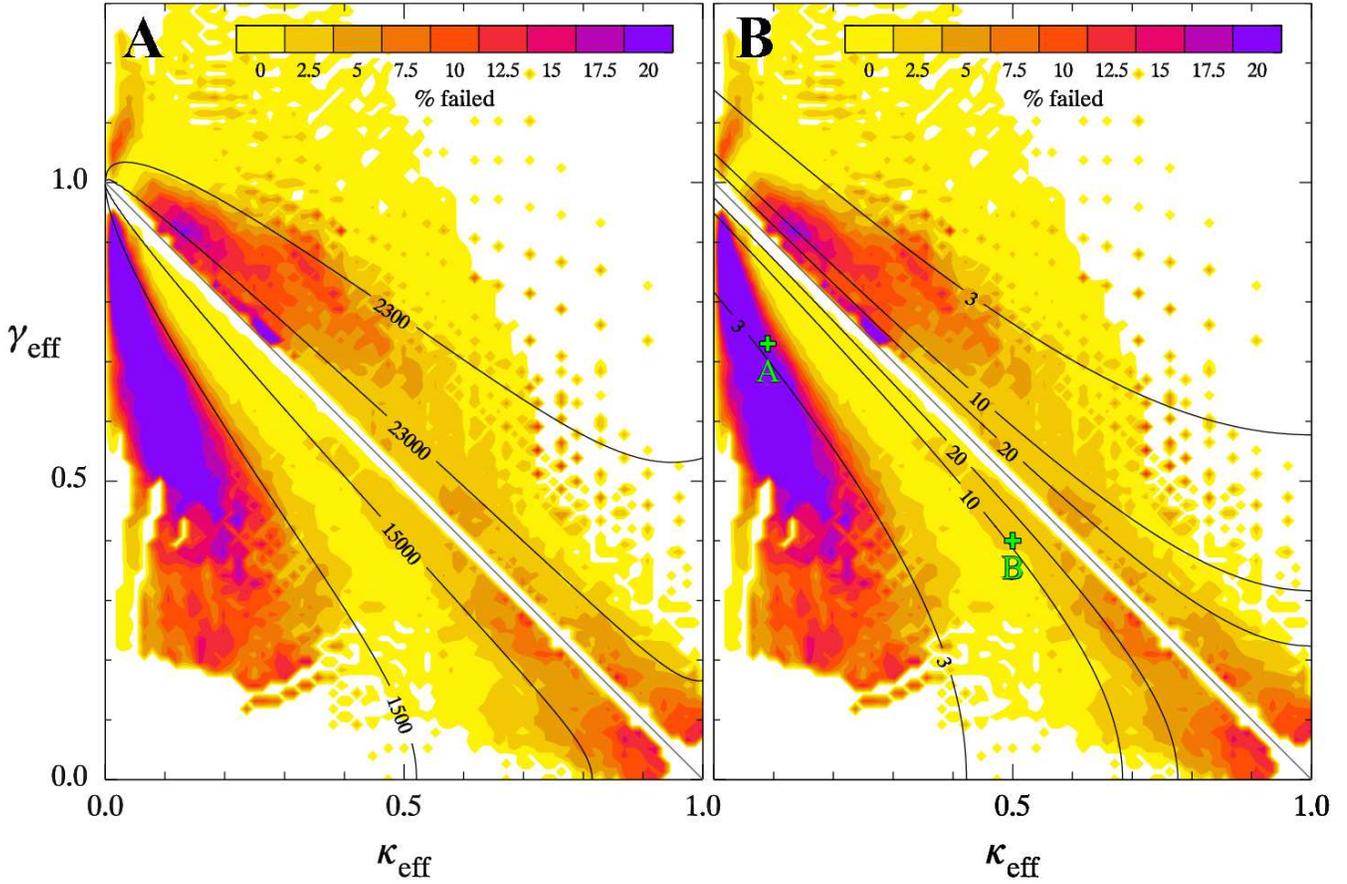}
\caption{Same as panel D of Fig. \ref{fig:eff_pairs}, with overplotted contours of $N_*$ (panel A) and $\mu_{\rm th}$ (panel B). In panel B we show two locations, A and B, around which the map MPDs are compared (Section \ref{sec:micro}) and light curves are extracted (\ref{sec:lcurves}). \label{fig:micro_comp}}
\end{figure*}

Panels A and B of Fig. \ref{fig:micro_comp} are the same as panel D of Fig. \ref{fig:eff_pairs}, with overplotted contours of $\mu_{\rm th}$ (equation \ref{eq:mu_th}) and the number of microlenses:
\begin{equation}
N_* = \frac{\kappa_{\rm eff}A}{\pi \langle M \rangle} \, ,
\end{equation}
for $s=0$, where $\langle M \rangle$ is the mean mass of the microlenses and $A$ is the area where they are randomly distributed.
The area $A$ is loosely defined and conventions are adopted according to which implementation of the ray--shooting technique is used.
In general, $A$ has to be larger than the relative size of the ray--shooting area in the lens plane and the receiving area in the source plane \citep[see section 2 of][for a discussion on how $A$ is defined in the {\tt GPU-D} code]{Bate2010}.

For our particular choice of $N_*$, the M1 region (panel A of Fig. \ref{fig:micro_comp}) is described accurately by $N_* < 1500$ for $\kappa_{\rm eff} < 0.3$, and similarly, the S1 region by $N_* > 3000$ for $\kappa_{\rm eff} < 0.3$.
For regions M2 and S2, it seems that $N_* >$ 15000 and 23000 respectively, but also $\mu_{\rm th} > 20$.
In general, the shapes of the MPDs in M1, M2, S1, and S2 seem to depend more strongly on the way different numbers of microlenses combine non-linearly through the lens equation \citep[e.g. equation 1 of][]{Vernardos2013} to create caustic networks.

\begin{figure}
\includegraphics[width=0.47\textwidth]{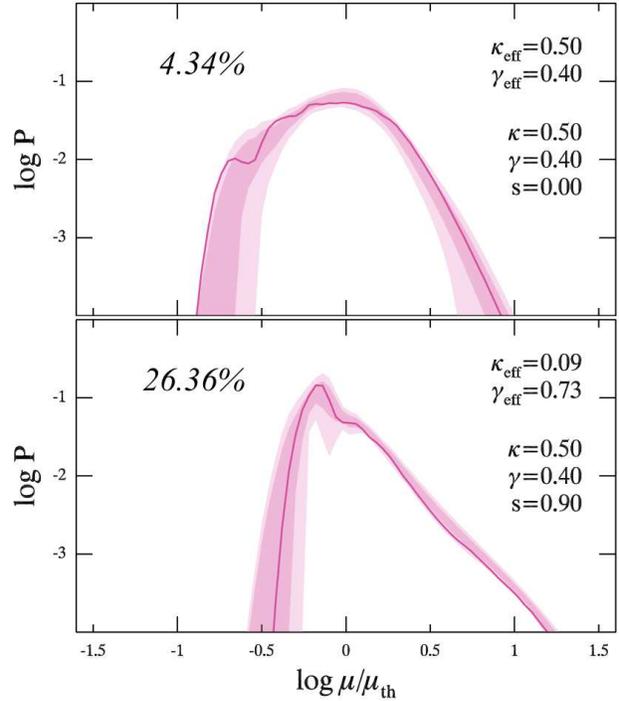}
\caption{1 and 2 standard deviation regions (shown in magenta) among a sample of 49 MPDs located within $r=0.04$ from $\kappa_0,\gamma_0 = (0.5,0.4)$. The percentages of failed tests are shown on each panel. The top panel is for $s=0.0$ and the bottom one for $s=0.9$, corresponding to locations B and A in panel B of Figure \ref{fig:micro_comp}. Location A lies inside the M1 region of panel D of Fig. \ref{fig:eff_pairs} and therefore has larger deviations and a higher percentage of failed KS tests between pairs of MPDs. The central MPD for $\kappa_0,\gamma_0$ is also shown (red line). Differences near the mode of the distributions are more significant, for example, although the central MPD in the top panel has a shape that deviates from the rest for $\mu < 0.3 \mu_{\rm th}$ the percentage of failed KS tests is lower than the bottom panel.\label{fig:mpds}}
\end{figure}

In Fig. \ref{fig:mpds} we show the $\pm 1$ and $\pm 2$ standard deviations among the shapes of 49 MPDs, which are within $r=0.04$, for two different locations in the minima region.
In both cases the $\kappa,\gamma$ values are the same, but $s$ is different, with one area lying inside M1 and the other outside (locations A and B in panel B of Figure \ref{fig:micro_comp}).
Differences around the peak of the distributions are the ones primarily responsible for the increased percentage of failed pairs (bottom panel of Fig. \ref{fig:mpds}), due to both the logarithmic scale used and the sensitivity of the KS test.
Discrepancies between the shapes of the MPDs that occur away from the peak have a lower probability and therefore do not effect the result of the KS test (top panel of Fig. \ref{fig:mpds}).

\subsection{Effect on macromodelling}
From the macromodel point of view, any derived $\kappa,\gamma$ values should be accompanied by $\Delta\kappa,\Delta\gamma$ uncertainties.
Our results indicate that depending on the location in the $\kappa,\gamma,s$, or equivalently $\kappa_{\rm eff},\gamma_{\rm eff}$, parameter space, such uncertainties may lead to statistically different MPDs of microlensing magnification maps.
We identify four such locations, M1, M2, S1, and S2, which are most clearly seen in Fig. \ref{fig:eff_pairs} for $r = $ 0.03 and 0.04.
For the radius $r$ of a circular area in the $\kappa,\gamma$ plane we have:
\begin{equation}
r = \sqrt{ \Delta\kappa^2 + \Delta\gamma^2 } \, .
\end{equation}
Therefore, $r=0.04$ corresponds to $\Delta\kappa,\Delta\gamma \approx 0.03$, assuming the same uncertainty in $\kappa,\gamma$, which can be considered a typical value (see Section \ref{sec:intro}).

\begin{figure*}
\includegraphics[width=\textwidth]{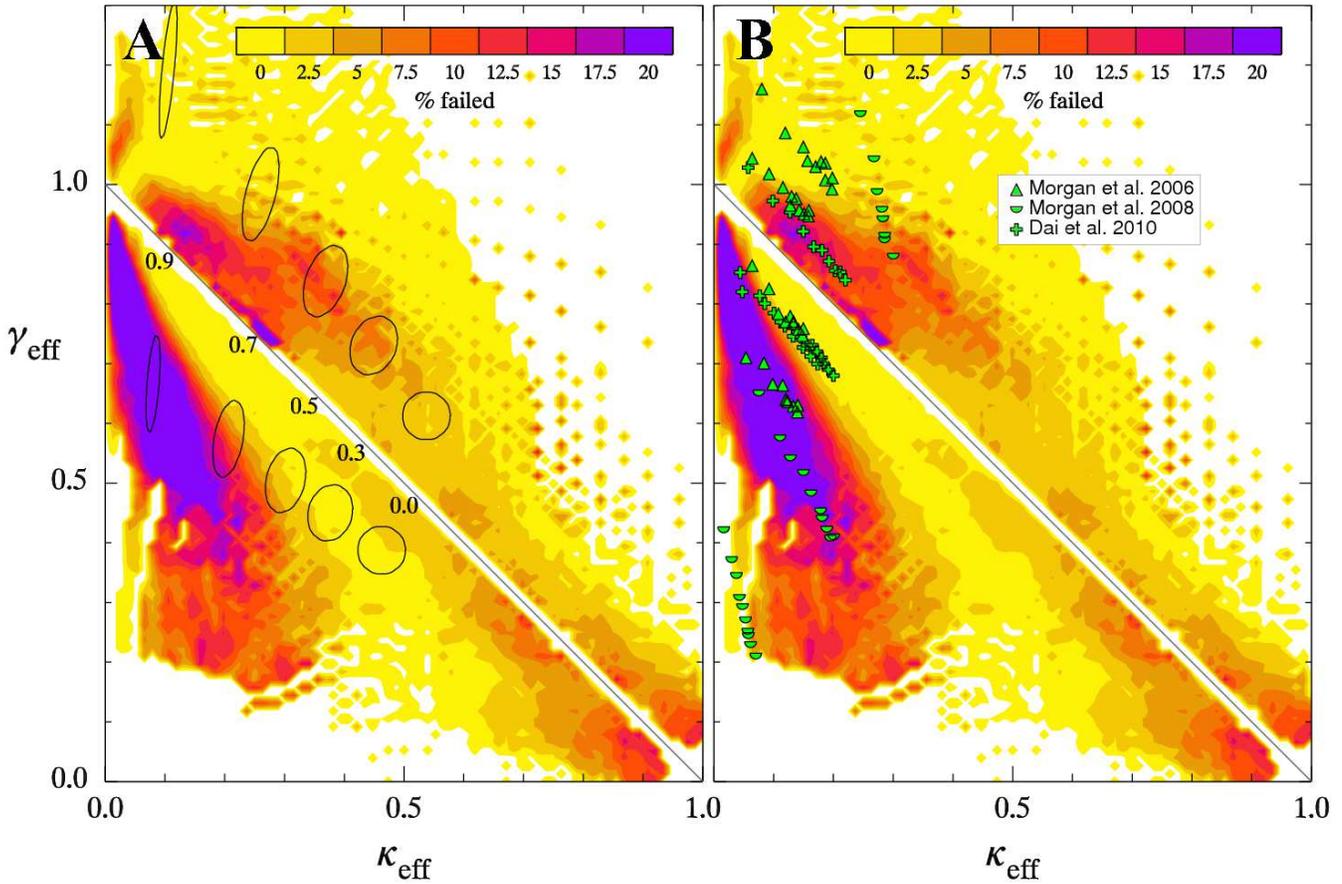}
\caption{Same as panel D of Fig. \ref{fig:eff_pairs}. The most probable $\kappa,\gamma$ values from a SIS macromodel are shown in panel A as black ellipses for different values of $s$. $\kappa,\gamma,s$ values for four gravitational lenses modeled with the de Vaucouleurs$+$dark matter halo mass profile are shown in panel B (green data points).\label{fig:macro_comp}}
\end{figure*}

\cite{Mediavilla2009} used SIS models for 20 systems, finding the most probable values of $\kappa$ to be 0.45 in the minima and 0.55 in the saddle--point region, while \cite{Schechter2002} used 0.475 and 0.525 as typical values.
\cite{Witt1995} showed that $\gamma = 3\kappa - 1$ holds at the position of the macroimages for this model.
Using this information we can assume that $\kappa,\gamma$ values from a SIS model will most likely lie within a circular area on the parameter space, centered halfway between the values of \cite{Schechter2002} and \cite{Mediavilla2009} and with a diameter equal to their separation.
Because $s$ is treated as a free parameter in the case of the SIS model, this circular area will transform according to equation (\ref{eq:eff}) in the $\kappa_{\rm eff},\gamma_{\rm eff}$ parameter space.
Examples for a few values of $s$ are shown in panel A of Fig. \ref{fig:macro_comp}.

For the SIS model, the most likely derived $\kappa,\gamma$ values are located in a region of parameter space where an uncertainty of $\Delta\kappa,\Delta\gamma = 0.03 - 0.04$ leads to magnification maps with statistically equivalent MPDs (see panel A of Fig. \ref{fig:macro_comp}).
However, this is true only for specific values of $s$ i.e. $s \le 0.5$ in the minima, $s < 0.3$ and $s \ge 0.7$ in the saddle--point region.
For the remaining values of $s$, an uncertainty of $\Delta\kappa,\Delta\gamma \le 0.02 - 0.01$ may be required.
This includes $s=0.9$, the most likely value for $s$ for a few systems \citep[e.g.][]{Mediavilla2009,Bate2011,Pooley2012}.
In such cases, accurate macromodels with well-constrained $\kappa,\gamma$ values should be preferred.

Another macromodelling approach was used by \cite{Dai2010} and \cite{Morgan2006,Morgan2008}, which constrains $s$ simultaneously with the values of $\kappa,\gamma$.
The values calculated by these authors for 4 multiply imaged systems are plotted in panel B of Fig. \ref{fig:macro_comp}.
The de Vaucouleurs$+$dark matter halo profiles assume a less cuspy mass distribution for the galaxy lens than the SIS, leading to lower values of $\kappa$.
In panel B of Fig. \ref{fig:macro_comp}, it can be seen that for most of the $\kappa,\gamma,s$ values from three studies using this macromodelling approach an uncertainty of $\Delta\kappa,\Delta\gamma \lesssim 0.02$ would be required to lead to statistically equivalent MPDs.
We note here that these studies generate simulated light--curves from convolved magnification maps rather than directly using the MPD.

\subsection{Effect on accretion disc constraints: MPDs}
Our results indicate that there are differences between the MPDs within areas of fixed $\Delta\kappa,\Delta\gamma$ in the parameter space.
For example, the MPDs in the bottom panel of Fig. \ref{fig:mpds} span almost an order of magnitude for $\mu = 0.8 \mu_{\rm th}$.
However, the impact of such uncertainties on derived accretion disc constraints is something to be determined by follow--up studies.

Let us assume that two multiply imaged quasars have the same underlying accretion disc, and the macromodels of their corresponding lensing galaxies provide $\kappa,\gamma$ values with negligible uncertainties.
If those $\kappa,\gamma$ values fall in an area of maps with statistically equivalent MPDs in the parameter space, then their microlensing observable properties are more likely to be the same.
Reversing this argument, if the observed properties of two microlensed quasars, whose macromodel $\kappa,\gamma$ values are accurately known and lie within a region of maps with statistically equivalent MPDs in parameter space, are found to be different, then the underlying accretion disc model is more likely to be different as well.
Although such scenarios may be presently unrealistic due to uncertainties effecting the microlensing observations and models (e.g. time delays, milli--lensing, macromodels, etc), this may change in the near future, after hundreds, or even thousands, of microlensed quasars will have been discovered.

\subsection{Effect on accretion disc constraints: light curves}
\label{sec:lcurves}
Another way of comparing maps, apart from the MPDs, is through extracted light curves.
Due to the large amount of additional computations required to perform such comparisons, we are comparing neighbouring maps from two locations in the parameter space.
Two trial values of $\kappa,\gamma$ have been selected, namely (0.53,0.43) and (0.47,0.37), located within $\Delta\kappa,\Delta\gamma = 0.03$ ($r \approx 0.04$) from a central value $\kappa_0,\gamma_0 = (0.5,0.4)$.
For $s=0.9$ we are in location A (see panel B of Figure \ref{fig:micro_comp}), where the KS test between the MPDs of neighbouring maps tends to fail, while for $s=0$ we are in location B, where the opposite is expected.

We assume a finite source size with which we are convolving the maps before extracting the light curves.
Thin-disc theory \citep{Shakura1973} gives the following relation for the radius of the accretion disc:
\begin{equation}
\label{eq:rtheory}
R = 9.7 \times 10^{15} \left( \frac{M}{10^{9} M_{\odot}} \right)^{\frac{2}{3}} \left( \frac{f_{\rm Edd}}{\eta} \right)^{\frac{1}{3}} \left( \frac{\lambda}{\mu m} \right)^{\frac{4}{3}} \quad \mathrm{cm},
\end{equation}
where we have assumed that the inner edge of the disc has a negligible effect.
Using typical values for the physical quantities appearing in this equation, viz. $10^{9}$M$_{\odot}$ for the mass of the supermassive black hole, 0.25 for the Eddington luminosity \citep[$f_{\rm Edd}$;][]{Kollmeier2006,Pooley2007,Blackburne2011}, and 0.15 for the accretion efficiency \citep[$\eta$;][]{Yu2002,Blackburne2011}, we end up with $\approx 3 \times 10^{15}$cm for the radius of the disc, as it is seen in the $u'$ ultraviolet filter ($354 \pm 31$ nm).
The mean value for the $R_{\rm Ein}$ (equation \ref{eq:Rein}) of 87 multiply imaged quasars is $5.11 \times 10^{16}$cm \citep[][for 1 M$_{\odot}$ microlenses]{Mosquera2011b}, which is the one that will be used in the following.
Therefore, the radius of a thin-disc given by equation (\ref{eq:rtheory}) as seen in the $u'$ filter would be $\approx$ 0.059 $R_{\rm Ein}$.

We approximate the brightness profile of a thin-disc at a given wavelength by a two-dimensional Gaussian distribution, having the value of the standard deviation, $\sigma$, set to match the radius of equation (\ref{eq:rtheory}).
We note here that the half--light radius of the disc should be used \citep[][]{Mortonson2005} if the goal would be to constrain the thin-disc model, which is not the case here.
We truncate the profile at a radius equal to $3\sigma$, which contains 99.7 per cent of the total profile brightness, and end up with a profile that has a diameter, or size, of $1.8 \times 10^{16}$cm, or 0.35 $R_{\rm Ein}$.
The GERLUMPH magnification maps that we are using have a width of 25 $R_{\rm Ein}$ and a resolution of 10000 pixels, meaning that our profile size would correspond to 142 map pixels.
We can now perform the convolution between the map and the profile to get the magnification map for a finite source with the chosen profile.

\begin{figure}
\includegraphics[width=0.47\textwidth]{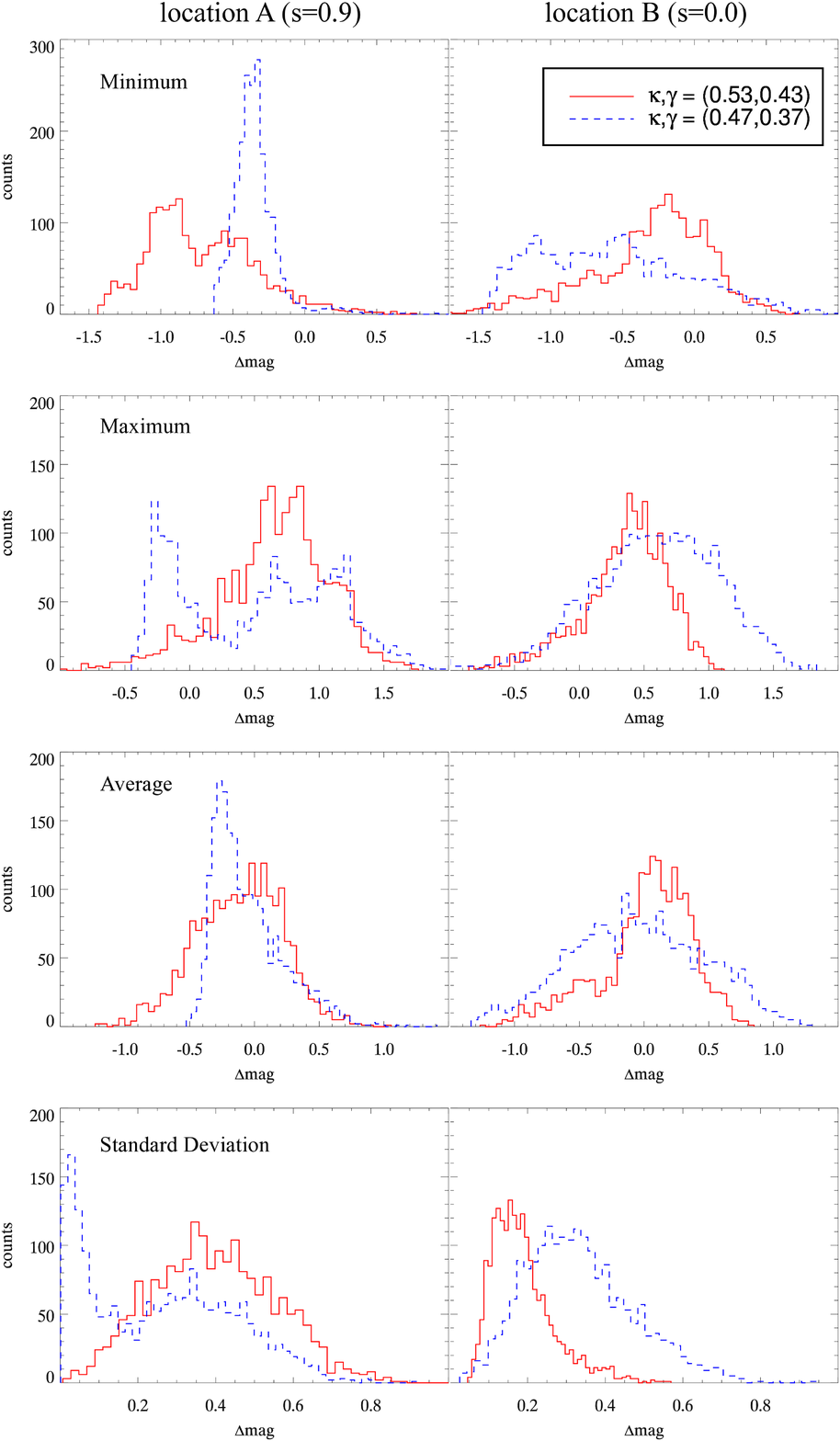}
\caption{Histograms of statistical properties of 2000 light curves extracted from neighbouring maps in two locations in parameter space. The magnification along the light curves is converted to magnitude change (equation \ref{eq:mu_Dm}) and basic statistical properties are calculated from each light curve, viz. the minimum, maximum, average, and standard deviation. Maps with $\kappa,\gamma = (0.53,0.43)$ are shown in solid (red) lines, and maps with $\kappa,\gamma = (0.47,0.37)$ are shown in dashed (blue) lines. Maps from location A (see panel B of Figure \ref{fig:micro_comp}) have $s=0.9$ and are shown on the left--hand side panels, while maps from location B have $s=0.0$ and are shown on the right--hand side.
\label{fig:lcurves}}
\end{figure}

A number of 2000 trial light curves has been randomly extracted from the maps and compared using basic statistical properties.
The length of the light curves has been set to $1 R_{\rm Ein}$ and continuous pixel sampling was assumed.
For each light curve we have calculated the minimum, maximum, average, and standard deviation, of the magnification values, and converted to magnitude change with respect to the macro--magnification (equation \ref{eq:mu_th}):
\begin{equation}
\label{eq:mu_Dm}
\Delta {\rm mag} = 2.5 \, \mathrm{log} ( \mu / \mu_{\rm th} ).
\end{equation}
In Figure \ref{fig:lcurves} we show the histogram of each statistical property for the two neighbouring maps in locations A and B; each row of panels shows the same statistical property and each column the same location in parameter space.

It is generally expected that the light curve properties will be not be exactly the same between the neighbouring maps from the same location.
The average $\Delta$mag of the light curves is expected to wash out any differences; the magnification is expected to vary around $\mu_{\rm th}$ (equation \ref{eq:mu_th}), which is very close for the two neighbouring maps at each location.
However, the distributions of the minimum, the maximum, and the standard deviation, are distinctly different in the left--hand panels of Figure \ref{fig:lcurves} (location A), as opposed to the right--hand ones (location B).
We choose to quantify this difference using the difference between the modes of the distributions, shown in Table \ref{tab:difmodes}.
Data in this table support the fact that light curves from the two maps in location A have more different statistical properties than those from location B.
We reach the same conclusion if we increase the number of extracted light curves, the length of the light curves, and the effect of randomly sampling the magnification map.
Finally, by reducing the distance in parameter space, i.e. choosing maps with $\Delta\kappa,\Delta\gamma = \pm0.02$ and $\pm$0.01 from $\kappa_0,\gamma_0$, we expect the light curve properties to converge.
However, we find that this happens faster for maps in location B,  while the differences between maps in location A persist.

The conclusions we have drawn based on our results from comparing MPDs using KS tests appear to be also valid when comparing statistical properties of light curves.
This happens at least in the two representative locations that we have tested, while the behaviour in the remaining of parameter space could be easily examined.
Further investigation of the light curve properties and their distribution in different locations in parameter space are out of the scope of this paper.

\begin{table}
\caption{Difference between the modes of the distributions shown in the left (location A) and right (location B) panels of Figure \ref{fig:lcurves}, in units of $\Delta$mag. If the mode of a given distribution is not unique (e.g. the distribution of the maxima for one of the maps in location A, shown as red line in left panel of Figure \ref{fig:lcurves}), then the average between the modes is used.}
\begin{center}
\begin{tabular}{|l|l|l|l|}
Location            & A            & B          \\
\hline
Minimum 			& 0.55         & 0.30       \\
Maximum				& 1.01         & 0.33       \\
Average 			& 0.30         & 0.21       \\
Standard deviation 	& 0.32         & 0.09       \\
\hline
\end{tabular}
\end{center}
\label{tab:difmodes}
\end{table}

\section{Conclusions}
We have used 55,900 microlensing magnification maps from the GERLUMPH online resource, which consists of a total of $\sim$70,000 maps, corresponding MPDs, and other supporting data.
We have compared the equivalence of neighbouring macromodels with $\Delta\kappa,\Delta\gamma$ uncertainties throughout the $\kappa,\gamma,s$ parameter space using the KS test on the MPDs.
The robustness of the KS test with respect to parameter space studies of thousands of MPDs has been demonstrated elsewhere \citep[][]{Vernardos2013,Vernardos2014a}.

We find that macromodel uncertainties of $\Delta\kappa,\Delta\gamma > 0.02$ can lead to significant differences between MPDs in certain areas of parameter space, which could potentially affect derived accretion disc model constraints.
However, the magnitude of such systematic errors is left to be determined by follow--up studies.
Such studies would be more well-suited for single, or a few, systems rather than the entire parameter space, making use of all the caustic network information included in a map, as opposed to using just the MPD.

Throughout the parameter space studied, microlensing maps within $\Delta\kappa,\Delta\gamma = 0.01$ from a fiducial $\kappa,\gamma$ value have statistically equivalent MPDs.
The exceptions are small areas close to the critical line, $\lvert 1-\kappa -\gamma \rvert \le 0.02$, for $\kappa \le 0.3$, $\kappa \ge 0.6$ and $s < 0.1$ (or $\kappa_{\rm eff} \le 0.3$, $\kappa_{\rm eff} \ge 0.6$; see Fig. \ref{fig:eff_pairs} for $r=0.01$).
This suggests that macromodel-derived $\kappa,\gamma$ values for the majority of known multiply imaged quasars need only be accurate to $\Delta\kappa,\Delta\gamma = 0.01$.

However, the impact of the uncertainties $\Delta\kappa,\Delta\gamma$ is bigger than the fiducial $\kappa,\gamma$ value they correspond to, and depends on the location in the parameter space.
There are large areas where values of $\Delta\kappa,\Delta\gamma =$ 0.03 or 0.04 lead to maps with statistically equivalent MPDs, described in Section \ref{sec:discussion} and shown in Fig. \ref{fig:eff_pairs}.
On the other hand, other areas of parameter space (M1,M2,S1 and S2 in Fig. \ref{fig:eff_pairs}) require $\Delta\kappa,\Delta\gamma \lesssim 0.02$ for statistical equivalence to hold.
Calculated $\kappa,\gamma,s$ values from the two most popular macromodelling approaches, namely, the de Vaucouleurs$+$dark matter halo and SIS mass profiles for the galaxy lens, tend to lie across both areas of parameter space.

Different strategies should be followed in producing and using microlensing magnification maps, based on parameter space location and macromodel uncertainty.
In the first case i.e. lying in an area of maps with statistically equivalent MPDs, any single map within this area can be used in convolutions with realistic accretion disc profiles to simulate observations.
Because a single magnification map covers a finite area of the source plane, sampling observational properties for large source profiles compared to the map dimensions could lead to results effected by small number statistics.
A common solution to this issue is to generate multiple maps for the same $\kappa,\gamma$ \citep[e.g.][]{Bate2008,Floyd2009}.
We assert here that an alternative solution is to use maps within a $\Delta\kappa,\Delta\gamma$ region of parameter space that have statistically equivalent MPDs.

In regions of parameter space where $\Delta\kappa,\Delta\gamma$ lead to maps with statistically different MPDs, a potential systematic error is introduced in the derived accretion disc constraints.
Such errors could be controlled by using all the maps within the $\Delta\kappa,\Delta\gamma$ area in combination with bootstrapping, or other statistical techniques.
If, however, areas of the source plane larger than a single map are required to increase the sample size, the computationally demanding task of generating series of maps with the same $\kappa,\gamma$ values (or generating larger maps to begin with) cannot be avoided.

Finally, we point out that $>$ 70,000 GERLUMPH maps are freely available to be used in any of the ways described above.
Thus, the computationally demanding task of generating hundreds of maps to model a specific system could be significantly reduced, or even become unnecessary.

\section*{acknowledgements}
This research was undertaken with the assistance of resources provided at gSTAR through the ASTAC scheme supported by the Australian Government.
gSTAR is funded by Swinburne and the Australian Government's Education Investment Fund.
The authors would like to thank Nick Bate for his comments on the final version of the paper.
We also thank the anonymous referee for comments that improved the clarity of our paper.

\bibliography{dkdg}

\label{lastpage}
\end{document}